\documentclass[aps,prl,twocolumn,groupedaddress,showpacs]{revtex4}

\usepackage{bm}
\usepackage{graphicx}
\usepackage{color}

\begin{document}

\title{Dimensionality reduction and band quantization induced by potassium intercalation in 1$T$-HfTe$_2$}

\author{Y. Nakata,$^1$ K. Sugawara,$^{1,2,3}$ A. Chainani,$^4$ K. Yamauchi,$^{5}$ K. Nakayama,$^{1}$ S. Souma,$^{2,3}$ P.-Y. Chuang,$^4$ C.-M. Cheng,$^4$ T. Oguchi,$^5$ K. Ueno,$^6$ T. Takahashi,$^{1,2,3}$ and T. Sato$^{1,2,3}$}

\affiliation{$^1$ Department of Physics, Tohoku University, Sendai 980-8578, Japan,}
\affiliation{$^2$ WPI-Advanced Institute for Materials Research (WPI-AIMR), Tohoku University, Sendai 980-8577, Japan,}
\affiliation{$^3$ Center for Spintronics Research Network (CSRN), Tohoku University, Sendai 980-8577, Japan,}
\affiliation{$^4$ National Synchrotron Radiation Research Center, Hshinchu 30077, Taiwan ROC,}
\affiliation{$^5$ Institute of Scientific and Industrial Research, Osaka University, Ibaraki, Osaka 567-0047, Japan,}
\affiliation{$^6$ Department of Chemistry, Saitama University, Saitama 338-8570, Japan}
\date{\today}

\begin{abstract}
We have performed angle-resolved photoemission spectroscopy on transition-metal dichalcogenide 1$T$-HfTe$_2$ to elucidate the evolution of electronic states upon potassium (K) deposition. In pristine HfTe$_2$, an in-plane hole pocket and electron pockets are observed at the Brillouin-zone center and corner, respectively, indicating the semimetallic nature of bulk HfTe$_2$, with dispersion perpendicular to the plane. In contrast, the band structure of heavily K-dosed HfTe$_2$ is obviously different from that of bulk, and resembles the band structure calculated for monolayer HfTe$_2$. It was also observed that lightly K-dosed HfTe$_2$ is characterized by quantized bands originating from bilayer and trilayer HfTe$_2$, indicative of staging. The results suggest that the dimensionality-crossover from 3D (dimensional) to 2D electronic states due to systematic K intercalation takes place via staging in a single sample. The study provides a new strategy for controlling the dimensionality and functionality of novel quantum materials.
\end{abstract}

\maketitle

Controlling the dimensionality of materials is one of the key challenges in condensed matter physics because dimensionality plays a crucial role in determining exotic physical properties and quantum phenomena \cite{Novoselov1,Klitzing,Cao,Zeng,Mak1,Mak2,Wang,Xi1,Lu,Xi2,Saito,Mak3}. It is often seen that reduction of dimensionality from 3D to 2D causes emergence of novel physical properties absent in 3D bulk, as represented by massless Dirac fermions in graphene \cite{Novoselov1}, quantum Hall effect in semiconductor heterostructures \cite{Klitzing}, valley-selective circular dichroism \cite{Cao,Mak1,Zeng}, and valley Hall effect \cite{Mak2}. Moreover, the 3D-to-2D crossover sometimes leads to more gigantic physical properties, as exemplified by a drastic increase in the superconducting transition temperature in monolayer FeSe \cite{Wang}, the enhancement of charge-density-wave (CDW) transition temperature \cite{Xi1} and upper-critical field \cite{Lu,Xi2,Saito} in transition-metal dichalcogenides (TMDs). These drastic changes in physical properties upon reducing the dimensionality are known to be inherently linked to the modification of the electronic band structure such as the spin splitting caused by the inversion-symmetry breaking \cite{Zhu,Nakata1} and the indirect-to-direct transition nature of band gap due to the absence of 3D chemical bonding \cite{Mak3}. Also, 2D systems are very sensitive to external stimuli like strain, charge doping, and electric field, so that the engineering of band structure is more feasible, making 2D systems a promising platform to explore novel physical properties.

Given the importance of controlling the dimensionality, a next important issue is how to control it. A widely used approach to reduce the dimensionality of materials from 3D to 2D is to mechanically or chemically exfoliate the sample \cite{Novoselov2,Joesen} (top-down approach). This can lead to an exotic change in properties as highlighted by the realization of quantum Hall effect in graphene exfoliated from graphite \cite{Novoselov2}. A bottom-up approach such as deposition of atoms or molecules on a substrate using molecular-beam-epitaxy method \cite{Koma,Zhang,Nakata1,Nakata2,Sugawara} or chemical-vapor-deposition technique \cite{Sutter,Kim}, is also useful. While these approaches have been successfully employed to explore unconventional physical properties associated with the reduction of dimensionality, a systematic control of dimensionality in these approaches needs great efforts because it requires one-by-one fabrication of various films with different thickness. As reported here, we have found a rather unexpected and effective way to systematically control the dimensionality and band structure of a TMD system. 

In this Rapid Communication, we carried out a simple and useful approach to control the dimensionality and electronic structure on the surface of the TMD 1$T$-HfTe$_2$. Using angle-resolved photoemission spectroscopy (ARPES) for visualizing the band structure of 1$T$-HfTe$_2$, we first show that the pristine bulk compound is a typical semimetal with hole and electron pockets at the $\Gamma$ and M points in the Brillouin zone. Very surprisingly, our results also show that the original 3D electronic structure in bulk pristine 1$T$-HfTe$_2$ converts into a purely 2D electronic structure upon potassium (K) intercalation. Intriguingly, while the observed valence-band (VB) structure in pristine HfTe$_2$ is well reproduced by the band calculations for bulk, those in lightly and heavily K-dosed HfTe$_2$ well follow the calculated bands for $bilayer/trilayer$ and $monolayer$, respectively, providing evidence for dimensionality reduction and band quantization driven by K intercalation.

\begin{figure}
\includegraphics[width=1.65in, bb=130 60 370 600]{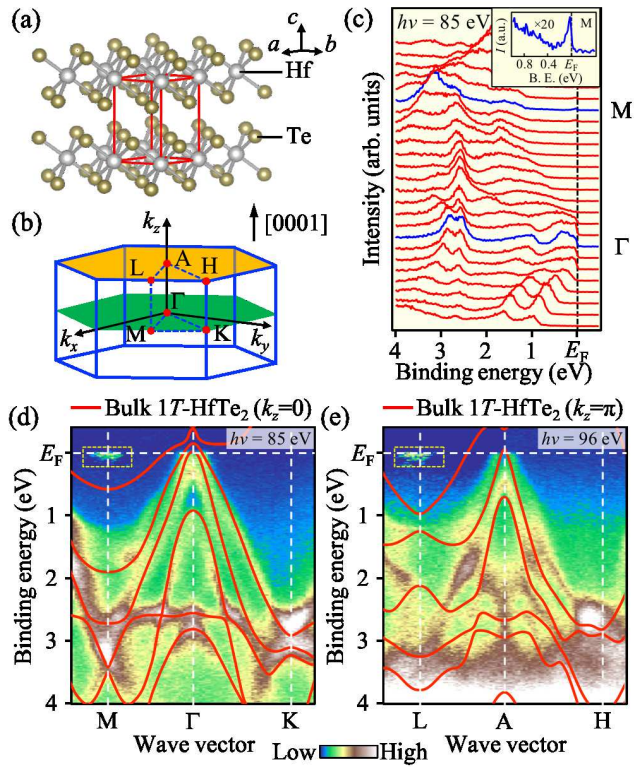}
	\vspace{1.0cm}
\caption{(color online): (a) Crystal structure of bulk 1$T$-HfTe$_2$. (b) Bulk hexagonal Brillouin zone (blue). Green and yellow hexagons correspond to the $k{_z}$=0 and $\pi$ planes, respectively. (c) EDCs of HfTe$_2$ measured along the $\Gamma$M cut with $h{\nu}$ = 85 eV. Inset shows an expansion of near-$E{\rm_F}$ EDC at the M point. (d) VB-ARPES intensity as a function of wave vector and $E{\rm_B}$, measured at $h{\nu}$ = 85 eV along the M$\Gamma$K cut ($k{_z}$=0 plane), together with the calculated band structure (red curves). Calculated bands are shifted downward by 470 meV to match the experimental VB. (e) Same as (d) but for the $k{_z}$=$\pi$ plane, obtained at $h{\nu}$ = 96 eV. The intensities enclosed by the dashed rectangle near $E{\rm_F}$ in (d) and (e) are shown with enhanced color contrast to better visualize the electron pocket.}
\end{figure}

We have chosen HfTe$_2$ as a test material to carefully monitor the evolution of electronic states upon K deposition, without complications from various orders (such as CDW and superconductivity) \cite{Brattas,Hodul1,Hodul2,Klipstein,Aminalragia-Giamini,Mangelsen} which are known to exist in many other TMDs. High-quality single crystals of 1$T$-HfTe$_2$ [see Fig. 1(a) for the crystal structure] were grown by the chemical-vapor-transport method \cite{Ueno}. ARPES measurements were performed at the beamlines BL28 of Photon Factory, KEK and BL21B1 of Taiwan Light Source, NSRRC. First-principles band-structure calculations were carried out by a projector augmented wave method \cite{VASP} with generalized gradient approximation (GGA) \cite{GGA}. For details of experiments and calculations, see sections 1-3 of Supplemental Material.

First, we show the electronic states of pristine HfTe$_2$. Figure 1(c) displays the energy distribution curves (EDCs) in the VB region measured along the $\Gamma$M cut in the bulk Brillouin zone (BZ) [Fig. 1(b)] at photon energy ($hv$) of 85 eV. We recognize several dispersive bands such as holelike bands centered at the $\Gamma$ point. One of these holelike bands crosses the Fermi level ($E_{\rm_F}$), forming a small hole pocket at the $\Gamma$ point. A low-intensity but clear Fermi-edge cut-off is seen around the M point [see the inset in Fig. 1(c) and Fig. 2(c)], indicating a small electron pocket around the M point. These results confirm the semimetallic nature of HfTe$_2$, consistent with previous transport measurements \cite{Klipstein,Mangelsen} and first-principles band-structure calculations \cite{Aminalragia-Giamini,Mangelsen}. To see the VB structure in more detail, we show in Figs. 1(d) and 1(e) the experimental band dispersions obtained by plotting the ARPES intensity as a function of wave vector ($k$) and binding energy ($E_{\rm_B}$) along high-symmetry cuts in the $\Gamma$KM ($k{_z}$ = 0) and AHL ($k{_z}$ = $\pi$) planes, respectively. We also show the corresponding band dispersions calculated for bulk 1$T$-HfTe$_2$ for comparison. As shown in Fig. 1(d), the overall experimental VB dispersion such as the location of dispersive holelike bands at the $\Gamma$ point is well reproduced in the calculation when the calculated bands are shifted downward as a whole by 0.47 eV. These holelike bands are attributed to the Te 5$p$ orbital. On the other hand, the electron pocket at the M point stemming from the Hf 5$d$ orbital \cite{Aminalragia-Giamini,Mangelsen} is significantly larger in the calculation. This is not due to the imperfect compensation of electrons and holes in the experiment, but due to the overestimation of the semimetallic band overlap ($i.e.$ negative band gap) in the calculations ($\sim$ 1 eV in contrast to 0.05 eV in experiments). Such overestimation was also recognized in previous band calculations of HfTe$_2$ \cite{Aminalragia-Giamini,Mangelsen} and also in other TMDs such as TiSe$_2$ \cite{Chen}. While one may attribute such overestimation to the unoptimized interlayer coupling, we found that the change in the $c$-axis lattice constant ($i.e.$ layer spacing) in the calculation is insufficient to correctly reproduce the experimental band structure, suggesting that there are other factors to overestimate the band gap in the calculation (see section 4 of Supplemental Material for details). As shown in Figs. 1(d) and 1(e), while the overall experimental VB dispersion looks similar between the $k{_z}$ = 0 and $\pi$ plane, a closer look reveals some characteristic differences such as the number of holelike bands and presence/absence of a flat dispersion at $E{\rm_B}$ $\sim$ 2.6 eV for $k{_z}$ = 0 / $k{_z}$ = $\pi$, indicating the 3D nature of the band structure. It is noted that the agreement of band structure between experiments and calculations is relatively poor in the $k{_z}$ = $\pi$ plane compared to that in the $k{_z}$ = 0 plane. In fact, it was necessary to shift the calculated VB dispersions upward by 0.28 eV to make a better match to the experimental data for $k{_z}$ = $\pi$. 

Now that the band structure of pristine HfTe$_2$ is established, next we demonstrate how the electronic states are influenced by K deposition. One would naturally expect that K atoms donate electrons into the HfTe$_2$ top layers.  This is clearly visible as a downward shift of the electronlike band at the M point, as shown in Figs. 2(a) and 2(b). We have estimated the energy shift ($E{\rm_{shift}}$) to be 165 meV by comparing the peak position of EDCs  as shown in Fig. 2(c). The electron doping with K deposition is also seen in the change of Fermi surface [Figs. 2(d) and 2(e)]. Upon K deposition, the hole pocket at the $\Gamma$ point disappears and at the same time the ellipsoidal electron pocket at the M point expands. Hereafter we label each sample with the size of electron pocket at the M point (See Section 5 of Supplemental Material for the detailed procedure), which corresponds to the electron density in a unit layer.  For example, samples in Figs. 2(d) and 2(e) are labeled $n{\rm_e}$=0.02 and $n{\rm_e}$=0.19 samples, respectively.

\begin{figure}
\includegraphics[width=1.55in, bb=140 60 370 380]{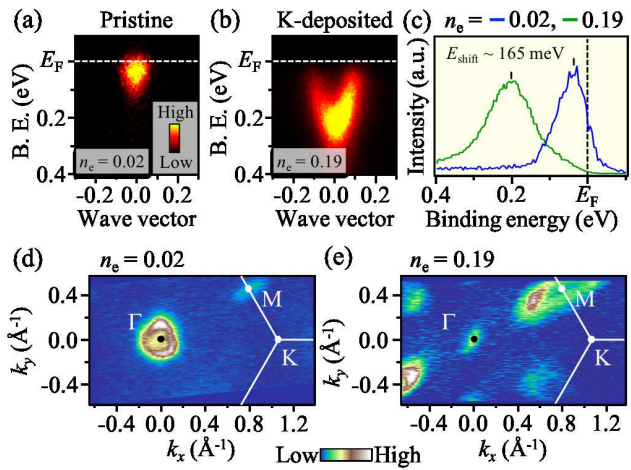}\vspace{1.0cm}
\caption{(color online): (a), (b) Near-$E{\rm_F}$ ARPES intensity along the KMK cut plotted as a function of wave vector and $E{\rm_B}$ for pristine ($n{\rm_e}$ = 0.02) and K-deposited ($n{\rm_e}$ = 0.19) samples, respectively. $n{\rm_e}$ represents the sheet electron concentration estimated from the size of the electron pocket at the M point. (c) EDC at the M point for $n{\rm_e}$ = 0.02 (blue curve) and 0.19 (green curve). (d), (e) ARPES-intensity mapping at $E{\rm_F}$ as a function of in-plane wave vector for $n{\rm_e}$ = 0.02 and 0.19, respectively. The intensity at $E{\rm_F}$ was obtained by integrating the ARPES-intensity within $\pm$20 meV of $E{\rm_F}$.}
\end{figure}

\begin{figure}
\centering
\includegraphics[width=1.6in, bb=130 60 370 870]{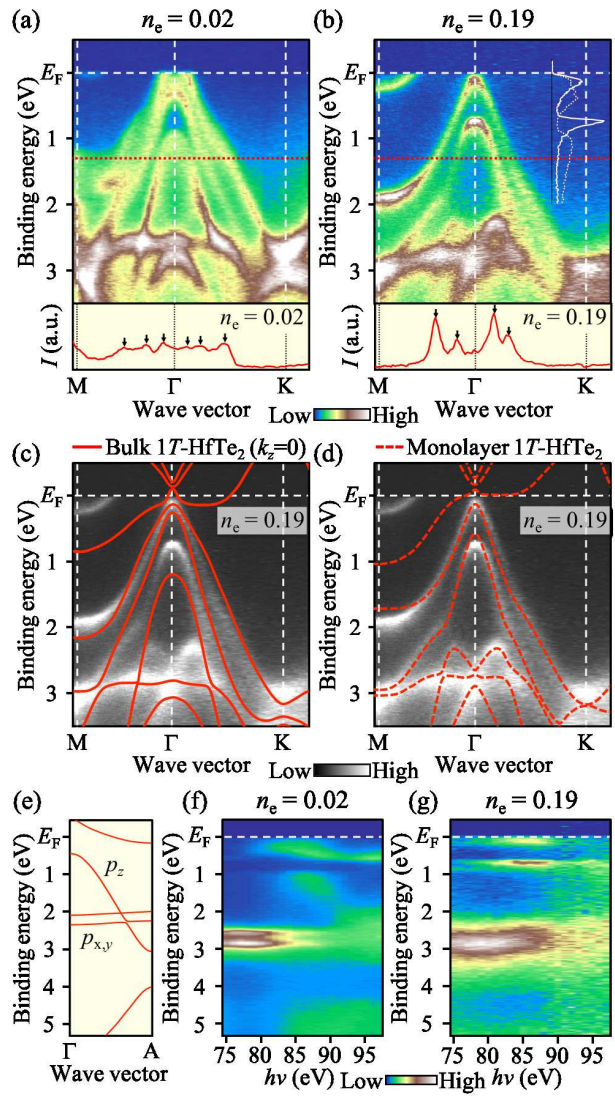}
\vspace{1.0cm}
\caption{(color online): (a), (b) VB-ARPES intensity along the M$\Gamma$K cut measured at $h{\nu}$ = 85 eV for $n{\rm_e}$ = 0.02 and 0.19, respectively. Bottom panel shows the MDC at $E{\rm_B}$ = 1.3 eV  (dashed line in the top panel), with $k$ position of bands indicated by arrows. Comparison of EDC at the $\Gamma$ point between $n{\rm_e}$ = 0.02 (dashed curve) and 0.19 (solid curve) is also shown in (b). (c), (d) First-principles band-structure calculations for bulk ($k{_z}$ = 0) and monolayer, respectively, compared with the ARPES intensity for $n{\rm_e}$ = 0.19 [same as (b) but plotted with gray scale]. Calculated bands were shifted downward by 700 and 925 meV, respectively. (e) Calculated band structure along the $\Gamma$A line for bulk HfTe$_2$. (f), (g) Normal-emission ARPES intensity as a function of $h{\nu}$ and $E{\rm_B}$ for $n{\rm_e}$ = 0.02 and 0.19, respectively.}
\end{figure}

\begin{figure*}
\includegraphics[width=2.15in, bb=340 60 640 410]{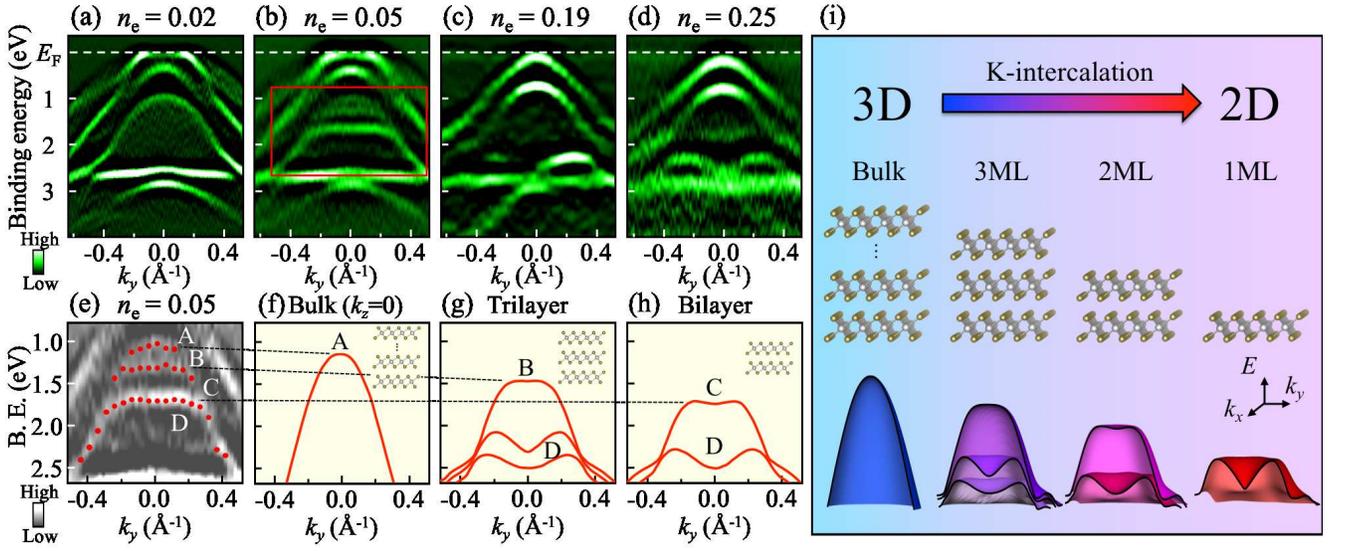}\vspace{1.0cm}
\caption{(color online): (a)-(d) Second-derivative of VB-ARPES intensity measured at  $h{\nu}$ = 85 eV for various samples with different K-deposition time, labeled by $n{\rm_e}$. (e) Expanded view of area enclosed by red rectangle in (b). Red dots are experimental band dispersions estimated from the peak position of EDCs. (f)-(h) Calculated band structure for bulk, trilayer, and bilayer HfTe$_2$, respectively. The holelike bands at lower $E{\rm_B}$ are not shown for clarity. (i) Schematic view of the evolution of band structure with K intercalation.}
\end{figure*}

Figures 3(a) and 3(b) show a side-by-side comparison of experimental band dispersions between pristine ($n{\rm_e}$=0.02) and heavily K-deposited ($n{\rm_e}$=0.19) sample. Although the electron pocket at the M point is shifted downward by K deposition as expected, the data show several anomalous changes in the band structure which cannot be  explained within a simple rigid-band scheme. For example, in the $n{\rm_e}$=0.02 sample, we observe three holelike bands centered at the $\Gamma$ point; two topped at around $E{\rm_F}$ and one topped at $\sim$ 1 eV. On the other hand, only two holelike bands (topped at $E{\rm_B}$ $\sim$ 0.1 and 0.75 eV) exist in the $n{\rm_e}$=0.19 sample. Such a difference in the number of bands is also highlighted by the representative momentum distribution curve (MDC) in the bottom panels of Figs. 3(a) and 3(b) which signify the presence of broad three peaks and sharp two peaks in the $\Gamma$M cut (also in the $\Gamma$K cut) for $n{\rm_e}$ = 0.02 and 0.19, respectively. As shown in Figs. 3(a) and 3(b), while a relatively flat band at $E{\rm_B}$ $\sim$ 2.6 eV appears to shift downward upon K deposition, a new M-shaped band emerges slightly above this band ($E{\rm_B}$ $\sim$ 2.5 eV) in the $n{\rm_e}$=0.19 sample. To clarify the origin of such anomalous variation of the band structure, we compare in Figs. 3(c) and 3(d) the experimental band structure for $n{\rm_e}$ = 0.19 with the calculated band structure for bulk and monolayer HfTe$_2$, respectively. One can see that the calculated band structure for monolayer shows a good agreement with the experimental band structure (except for the electron pocket at the M point), while the calculated band structure for the bulk apparently shows some disagreements such as in the location of the holelike bands and the absence of the M-shaped band. This suggests that the originally bulk-like band dispersion is ``converted" into the monolayer-like one upon K deposition. Such a change to the monolayer-like behavior is also confirmed by performing photon-energy-dependent ARPES measurements that signify a finite energy dispersion along $k{_z}$ in the pristine ($n{\rm_e}$=0.02) sample [Fig. 3(f)] in line with the band calculation [Fig. 3(e)], in contrast to no discernible $k{_z}$ dispersion in the  $n{\rm_e}$=0.19 sample [Fig. 3(g)]. These results indicate that the K deposition switches the dimensionality of electronic structure from 3D to 2D.  We found that such a 3D-2D transition is accompanied by a sharpening of the spectral line shape highlighted by the comparison of EDC at the $\Gamma$ point between  $n{\rm_e}$ = 0.02 and 0.19 in Fig. 3(b), which likely reflects absence of $k{_z}$-broadening effect and reduced contribution of photoelectron lifetime for $n{_e}$ = 0.19.

To gain further insight into the origin of observed dimensionality change, we investigated the evolution of electronic states for various K-deposition time and show the ARPES intensities for a series of $n{\rm_e}$ in Figs. 4(a)-4(d). This plot again confirms that the number of bands are apparently different between $n{\rm_e}$ = 0.02 and 0.19, reflecting the dimensionality change. One can also see that the band structures for $n{\rm_e}$ = 0.19 and 0.25 are very similar to each other. This suggests that the 2D nature of electronic states is more or less established at $n{\rm_e}$ = 0.19 and further K deposition simply leads to extra electron doping with a small constant shift of the overall band structure to higher binding energies. We comment here that, while a previous study for epitaxy-grown atomic-layer HfTe$_2$ thin film on AlN \cite{Aminalragia-Giamini} reported the Dirac-semimetal phase characterized by the Dirac-cone band at the $\Gamma$ point, the present result shows no upper Dirac-cone-like band for monolayer HfTe$_2$ [Fig. 4(d)] (see Section 5 of Supplemental Material for details). This difference may be attributed to a substrate induced effect in epitaxial atomic-layer HfTe$_2$ film on AlN compared to the present study.

Finally, we present another very important finding. We found that the experimental band structure for $n{\rm_e}$ = 0.05 shows unexpected behavior at energies away from $E{\rm_F}$. As highlighted by the area enclosed by red rectangle in Fig. 4(b), there exist three holelike bands topped at the $E{\rm_B}$ range of 1-1.7 eV. Such multiple bands are absent in other samples, and hence they are likely a characteristic of an intermediate state between 3D ($n{\rm_e}$ = 0.02) and 2D ($n{\rm_e}$ = 0.19 - 0.25). To obtain further insight into the origin of such subband feature, we compare the ARPES-derived band dispersion [Fig. 4(e)] with the calculations for multilayer HfTe$_2$ [Figs. 4(f), 4(g), and 4(h)].  One can see that band A is due to the bulk band since its shape and energy position are well reproduced by the calculation for bulk HfTe$_2$ as shown in Fig. 4(f). This assignment is also corroborated with the observation of a similar band in the pristine sample ($n{\rm_e}$ = 0.02) [Fig. 4(a)]. On the other hand, bands B and C show a reasonable agreement with the calculated topmost quantized bands for trilayer and bilayer HfTe$_2$, respectively. This implies that the surface of the $n{\rm_e}$ = 0.05 sample is inhomogeneous in terms of the K concentration, and different domains coexist at the surface. It is thus likely that we simultaneously detect three domains ($i.e.$, bulk, trilayer, and bilayer) in the ARPES data [Fig. 4(b)] for $n{\rm_e}$ = 0.05.  It is noted that a weak feature labeled D in Fig. 4(e) may be ascribed to a mixture of the second and/or third quantized $p_z$ orbital in trilayer and bilayer domains.

We discuss the origin of observed intriguing change in the band structure. We found no obvious change in the LEED (low-energy-electron-diffraction) pattern upon K deposition, in particular, regarding the location of the LEED spot. This suggests that the in-plane lattice parameter does not change and no surface reconstruction takes place (see section 7 of Supplemental Material for details). Thus, the observed  evolution of band structure upon K deposition in Figs. 2-4 is not ascribed to the structural modulation of the HfTe$_2$ layer itself. A plausible explanation for the observation of monolayer-like band dispersion for $n{\rm_e}$ = 0.19 and 0.25 is that K atoms are intercalated into the van der Waals gap of HfTe$_2$ layers around the surface, as naively understood by referring to stage-one graphite intercalation compounds (GICs) where atoms are intercalated in all the available van der Waals gaps. Since K atoms are randomly placed in HfTe$_2$ as inferred from the absence of band folding and additional LEED spots after K deposition, the K atoms do not enhance the interlayer coupling unlike the case of GICs where the periodic arrangement of intercalant atoms would promote the 3D nature of materials. In contrast, in the case of HfTe$_2$, each layer is effectively isolated from adjacent layers due to the increased layer spacing, leading to enhancement of monolayer-like nature. A similar behavior has been observed in K-intercalated MoS$_2$ \cite{Eknapakul1} and H-intercalated graphene on SiC \cite{H-SiC}. On the other hand, besides the monolayer-like feature, we found that a multiple staging from stage-two to stage-three takes place at the surface of a single sample and gives rise to emergence of several quantized bands in the lightly K-deposited regime ($n{\rm_e}$ = 0.05). This finding is of particular significance since we could experimentally demonstrate that the dimensionality of the electronic states (in other words, staging of the intercalation) around the surface can be $systematically$ and $easily$ controlled by the simple K-deposition technique.

The above quantization picture is further corroborated by considering the orbital character for the observed subbands. As visible in Fig. 4(b), the band quantization is well resolved only for the bands located at $\sim$1-2 eV. These bands originate from the $p{_z}$ orbital which is highly dispersive along the $k{_z}$ direction, as seen in Fig. 3(e). This situation is favorable for forming the quantized bands since the quantum confinement occurs along the $z$ direction (perpendicular to the surface). In contrast, the $p{_x}$ and $p{_y}$ orbitals are unlikely to be well quantized because of their weak $k{_z}$ dispersion; this is indeed inferred from the absence of subbands for the $p{_{x,y}}$-derived flat band at $\sim$2.6 eV in Figs. 4(a)-4(d) [see also Fig. 3(e)]. It is worthwhile to note here that the previous study reported similar subbands after Na intercalation in HfSe$_2$ \cite{Eknapakul2}. However, the mechanism of subband formation is totally different from the present case since the subbands of Na-intercalated HfSe$_2$ have a $p{_{x,y}}$ character and they originate from the in-plane lattice strain \cite{Eknapakul2}. This is also consistent with the ARPES and LEED data (see section 7 of Supplemental Material) of HfTe$_2$ which show no discernible variation of the in-plane lattice constant upon K intercalation in support of a weak strain effect. The present study demonstrates for the first time that the band quantization and dimensionality of electronic states can be manipulated by a simple deposition of atoms on the sample surface, as highlighted in Fig. 4(i).

We emphasize that the method proposed here to control the dimensionality and visualize the electronic states is useful since it can be performed on a single sample; this could be contrasted to the so-far established exfoliation and MBE techniques in which systematic control is rather difficult because they require one-by-one fabrication of various films with different thickness. Also, it is expected that our method can be widely applicable to other layered materials including TMDs if the condition of intercalation such as the species of alkali metals and evaporation temperature is optimized for each material (this point is important since the intercalation/adsorption condition would strongly depend on the combination of alkali-metal elements and constituent elements of TMDs \cite{Biswas, Kang}). While we selected HfTe$_2$ to monitor the evolution of electronic states to avoid complications from various orders \cite{Brattas,Hodul1,Hodul2,Klipstein,Aminalragia-Giamini,Mangelsen} and to effectively demonstrate controllability of dimensionality, a choice of other TMDs would provide us a precious opportunity to study in a systematic way the interplay between dimensionality and various exotic physical properties, such as unconventional superconductivity \cite{Lu,Xi2,Saito}, ferromagnetism \cite{Bonilla}, topological phase transition, and quantum spin Hall effect \cite{Qian,Tang,Fei,Wu}. Experiments combining surface spectroscopies and magneto-transport measurements in K-deposited TMDs would be highly desired in future.

In conclusion, our ARPES study on HfTe$_2$ revealed a rich variation of electronic structure upon K intercalation associated with the 3D-to-2D crossover. We proposed a new technique to control the dimensionality of electronic states at the surface by simple K deposition. The present result would serve as a foundation for investigating the interplay between dimensionality and exotic physical properties in TMDs and other layered quantum materials.

\begin{acknowledgments}
We thank T. Kato, K. Hori, K. Owada, T. Nakamura, H. Oinuma, K. Shigekawa, and D. Takane for their assistance in the ARPES measurements. We also thank KEK-PF for access to beamline BL28 (Proposal number: 2018S2-001) and NSRRC-TLS for beamline BL21A2. This work was supported by JST-CREST (No: JPMJCR18T1), MEXT of Japan (Innovative Area ``Topological Materials Science" JP15H05853), JSPS (JSPS KAKENHI No: JP17H01139, JP26287071, JP18H01160, JP18H01821, 18K18986, JP25107003, JP25107004, and 18J10038), Grant for Basic Science Research Projects from the Sumitomo Foundation, and Murata Science Foundation. Y. N. acknowledges support from GP-Spin at Tohoku University.
\end{acknowledgments}

\bibliographystyle{prsty}

\begin{thebibliography}{50}
\bibitem{Novoselov1} K. S. Novoselov, A. K. Geim, S. V. Morozov, D. Jiang, M. I. Katsnelson, I. V. Grigorieva, S. V. Dubonos, and A. A. Firsov, Nature {\bf 438}, 197 (2005).
\bibitem{Klitzing} K. V. Klitzing, G. Dorda, and M. Pepper, Phys. Rev. Lett. {\bf 45}, 494 (1980).
\bibitem{Cao} T. Cao, G. Wang, W. Han, H. Ye, C. Zhu, J. Shi, Q. Niu, P. Tan, E. Wang, B. Liu, and J. Feng, Nat. Commun. {\bf 3}, 887 (2012).
\bibitem{Zeng} H. Zeng, J. Dai, W. Yao, D. Xiao, and X. Cui, Nat. Nanotech. {\bf 7}, 490 (2012).
\bibitem{Mak1} K. F. Mak, K. He, J. Shan, and T. F. Heinz, Nat. Nanotech. {\bf 7}, 494 (2012).
\bibitem{Mak2} K. F. Mak, K. L. McGill, J. Park, and P. L. McEuen, Science {\bf 344}, 1489 (2014).
\bibitem{Wang} Q. Y. Wang, Z. Li, W. H. Zhang, Z. C. Zhang, J. S. Zhang, W. Li, H. Ding, Y. B. Ou, P. Deng, K. Chang, J. Wen, C. L. Song, K. He, J. F. Jia, S. H. Ji, Y. Y. Wang, L. L. Wang, X. Chen, X. C. Ma, and Q. K. Xue, Chin. Phys. Lett. {\bf 29}, 037402 (2012).
\bibitem{Xi1} X. Xi, L. Zhao, Z. Wang, H. Berger, L. Forr$\rm\acute{o}$, J. Shan, and K. F. Mak, Nat. Nanotechnol. {\bf 10}, 765 (2015).
\bibitem{Lu} J. M. Lu, O. Zheliuk, I. Leermakers, N. F. Q. Yuan, U. Zeitler, K. T. Law, and J. T. Ye, Science {\bf 350}, 1353 (2015).
\bibitem{Saito} Y. Saito, Y. Nakamura, M. S. Bahramy, Y. Kohama, J. Ye, Y. Kasahara, Y. Nakagawa, M. Onga, M. Tokunaga, T. Nojima, Y. Yanase, and Y. Iwasa, Nat. Phys. {\bf 12}, 144 (2015).
\bibitem{Xi2} X. Xi, Z. Wang, W. Zhao, J. -H. Park, K. T. Law, H. Berger, L. Forr$\rm\acute{o}$, J. Shan, and K. F. Mak, Nat. Phys. {\bf 12}, 139 (2016).
\bibitem{Mak3} K. F. Mak, C. Lee, J. Hone, J. Shan, and T. F. Heinz, Phys. Rev. Lett. {\bf 105}, 136805 (2010).
\bibitem{Zhu} Z. Y. Zhu, Y. C. Cheng, and U. Schwingenschl$\rm\ddot{o}$gl, Phys. Rev. B {\bf 84}, 153402 (2011).
\bibitem{Nakata1} Y. Nakata, K. Sugawara, S. Ichinokura, Y. Okada, T. Hitosugi, T. Koretsune, K. Ueno, S. Hasegawa, T. Takahashi, and T. Sato, npj 2D Mater. Appl. {\bf 2}, 12 (2018).
\bibitem{Novoselov2} K. S. Novoselov, D. Jiang, F. Schedin, T. J. Booth, V. V Khotkevich, S. V Morozov, and A. K. Geim, Proc. Natl. Acad. Sci. {\bf 102}, 10451 (2005).
\bibitem{Joesen}P. Joesen, R. F. Frindt, and S. R. Morrison, Mater. Res. Bull. {\bf 21}, 457 (1986).
\bibitem{Koma} A. Koma, K. Sunouchi, and T. Miyajima, Microelectron. Eng. {\bf 2}, 129 (1984).
\bibitem{Zhang} Y. Zhang, T. R. Chang, B. Zhou, Y. T. Cui, H. Yan, Z. Liu, F. Schmitt, J. Lee, R. Moore, Y. Chen, H. Lin, H. T. Jeng, S. K. Mo, Z. Hussain, A. Bansil, and Z. X. Shen, Nat. Nanotech. {\bf 9}, 111 (2014).
\bibitem{Nakata2} Y. Nakata, K. Sugawara, R. Shimizu, Y. Okada, P. Han, T. Hitosugi, K. Ueno, T. Sato, and T. Takahashi, NPG Asia Mater. {\bf 8}, e321 (2016). 
\bibitem{Sugawara} K. Sugawara, Y. Nakata, R. Shimizu, P. Han, T. Hitosugi, T. Sato, and T. Takahashi, ACS Nano {\bf 10}, 1341 (2016). 
\bibitem{Sutter} P. W. Sutter, J. I. Flege, and E. A. Sutter, Nat. Mater. {\bf 7}, 406 (2008).
\bibitem{Kim} D. Kim, D. Sun, W. Lu, Z. Cheng, Y. Zhu, D. Le, T. S. Rahman, and L. Bartels, Langmuir {\bf 27}, 11650 (2011).
\bibitem{Brattas} L. Brattas and A. Kjekshus, Acta Chem. Scand. {\bf 27}, 1290 (1973).
\bibitem{Hodul1} D. Hodul and M. J. Sienko, Physica {\bf 99B}, 215 (1980).
\bibitem{Hodul2} D. T. Hodul and A. M. Stacy, J. Phys. Chem. Solids {\bf 46}, 1447 (1985).
\bibitem{Klipstein} P. C. Klipstein, D. R. P. Guy, E. A. Marseglia, J. I. Meakin, R. H. Friend, and A. D. Yoffe, J. Phys. C: Solid State Phys. {\bf 19}, 4953 (1986).
\bibitem{Aminalragia-Giamini} S. Aminalragia-Giamini, J. Marquez-Velasco, P. Tsipas, D. Tsoutsou, G. Renaud, and A. Dimoulas, 2D Mater. {\bf 4}, 015001 (2017).
\bibitem{Mangelsen} S. Mangelsen, P. G. Naumov, O. I. Barkalov, S. A. Medvedev, W. Schnelle, M. Bobnar, S. Mankovsky, S. Polesya, C. N\"ather, H. Ebert, and W. Bensch, Phys. Rev. B {\bf 96}, 205148 (2017).
\bibitem{Ueno} K. Ueno, J. Phys. Soc. Jpn. {\bf 84}, 121015 (2015).
\bibitem{VASP} G. Kresse and J. Furthmuller, Phys. Rev. B {\bf 54}, 11169 (1996).
\bibitem{GGA} J. P. Perdew, K. Burke and M. Ernzerhof, Phys. Rev. Lett. {\bf 77}, 3865 (1996).
\bibitem{Chen} P. Chen, Y.-H. Chan, X.-Y. Fang, Y. Zhang, M.Y. Chou, S.-K. Mo, Z. Hussain, A.-V. Fedorov, and T.-C. Chiang, Nat. Commun. {\bf 6}, 8943 (2015).
\bibitem{Eknapakul1} T. Eknapakul, P. D. C. King, M. Asakawa, P. Buaphet, R. H. He, S. K. Mo, H. Takagi, K. M. Shen, F. Baumberger, T. Sasagawa, S. Jungthawan, and W. Meevasana, Nano Lett. {\bf 14}, 1312 (2014).
\bibitem{H-SiC} C. Riedl, C. Coletti, T. Iwasaki, A. A. Zakharov, and U. Starke, Phys. Rev. Lett. {\bf 103}, 246804 (2009).
\bibitem{Eknapakul2} T. Eknapakul, I. Fongkaew, S. Siriroj, W. Jindata, S. Chaiyachad, S. K. Mo, S. Thakur, L. Petaccia, H. Takagi, S. Limpijumnong, and W. Meevasana, Phys. Rev. B {\bf 97}, 201104(R) (2018).
\bibitem{Biswas} D. Biswas, Alex M. Ganose, R. Yano, J. M. Riley, L. Bawden, O. J. Clark, J. Feng, L. Collins-Mcintyre, M. T. Sajjad, W. Meevasana, T. K. Kim, M. Hoesch, J. E. Rault, T. Sasagawa, David O. Scanlon, and P. D. C. King, Phys. Rev. B {\bf 96}, 085205 (2017).
\bibitem{Kang} M. Kang, B. Kim, S. H. Ryu, S. W. Jung, J. Kim, L. Moreschini, C. Jozwiak, E. Rotenberg, A. Bostwick, and K. S. Kim, Nano Lett. {\bf 17}, 1610 (2017).
\bibitem{Bonilla} M. Bonilla, S. Kolekar, Y. Ma, H. C. Diaz, V. Kalappattil, R. Das, T. Eggers, H. R. Gutierrez, M. H. Phan, and M. Batzill, Nat. Nanotech. {\bf 13}, 289 (2018).
\bibitem{Qian} X. Qian, J. Liu, L. Fu, and J. Li, Science {\bf 346}, 1344 (2014).
\bibitem{Tang} S. Tang, C. Zhang, Di. Wong, Z. Pedramrazi, H. Z. Tsai, C. Jia, B. Moritz, M. Claassen, H. Ryu, S. Kahn, J. Jiang, H. Yan, M. Hashimoto, D. Lu, R. G. Moore, C. C. Hwang, C. Hwang, Z. Hussain, Y. Chen, M. M. Ugeda, Z. Liu, X. Xie, T. P. Devereaux, M. F. Crommie, S. K. Mo, and Z. X. Shen, Nat. Phys. {\bf 13}, 683 (2017).
\bibitem{Fei} Z. Fei, T. Palomaki, S. Wu, W. Zhao, X. Cai, B. Sun, P. Nguyen, J. Finney, X. Xu, and D. H. Cobden, Nat. Phys. {\bf 13}, 677 (2017).
\bibitem{Wu} S. Wu, V. Fatemi, Q. D. Gibson, K. Watanabe, T. Taniguchi, R. J. Cava, and P. Jarillo-Herrero, Science {\bf 359}, 76 (2018).



\end{thebibliography}

\end{document}